\documentclass[reprint,prb,amsmath,amssymb,aps,longbibliography]{revtex4-1}

\usepackage{graphicx}
\usepackage{dcolumn}
\usepackage{bm}
\usepackage{longtable}
\usepackage{array}

\begin{document}

\title{On the relationship between topological and geometric defects}

\author{Sin\'{e}ad M. Griffin}
\affiliation{Molecular Foundry, Lawrence Berkeley National Laboratory, Berkeley, CA 94720, USA}
\affiliation{Department of Physics, University of California Berkeley, Berkeley, CA 94720, USA}
\email{sgriffin@lbl.gov}
\homepage{http://sites.google.com/sineadv0}

\author{Nicola A. Spaldin}
\email{nicola.spaldin@mat.ethz.ch}
\homepage{http://www.theory.mat.ethz.ch}
\affiliation{Materials Theory, ETH Zurich,
Wolfgang-Pauli-Strasse 27, CH-8093 Zurich, Switzerland}

\begin{abstract}

The study of topology in solids is undergoing a renaissance following renewed interest in the properties of ferroic domain walls as well as recent discoveries regarding topological insulators and skyrmionic lattices. Each of these systems possess a property that is `protected' in a symmetry sense, and is defined 
rigorously using a branch of mathematics known as topology. In this article we review the formal definition of topological defects as they are classified in terms of homotopy theory, and discuss the precise symmetry-breaking conditions that lead to their formation. We distinguish topological defects from geometric defects, which arise from the details of the stacking or structure of the material but are not protected by symmetry. We provide simple material examples of both topological and geometric defect types, and discuss the implications of the classification on the resulting material properties. 
\end{abstract}

\maketitle

\newpage

\section{Introduction}
Topological defects in materials have a long and rich history in the context of domain walls separating equivalent low-symmetry states of different orientations in ferroics. Such defects are functional entities in their own right, often showing exotic behavior such as sheet superconductivity in the ferroelastic twin walls of tungsten oxide, WO$_3$\cite{Aird/Salje:1998} or fast ionic transport along twin walls in feldspar and perovskite structures\cite{Salje:2000}. 

Recently, interest in the concept of topology in condensed matter has exploded following the discovery of skyrmionic magnetic lattices\cite{Roessler/Bogdanov/Pfleiderer:2006} and the experimental verification of the existence topological insulators\cite{Konig_et_al:2007}. Like some ferroelastic and ferroelectric domain walls, both of these systems possess a property that is `protected' in a symmetry sense, and so can be rigorously defined by its corresponding mathematical topological characteristics. At the same time, there has been a flurry of exciting new discoveries regarding the properties of interfaces\cite{Ohtomo/Hwang:2004,Stengel/Vanderbilt/Spaldin_NatMat:2009}, domain walls\cite{Seidel_et_al:2009,Salje:2013}, vortices\cite{Yadav_et_al:2016}, and spin textures\cite{Spaldin_et_al:2013}, particularly in ferroic and multiferroic systems, which it is tempting, although not always strictly correct, to discuss within a topological framework.  

In this paper we review the distinction between topological defects as they are strictly classified in terms of mathematical homotopy theory, and other defects arising from stacking or structural aspects of a material for which the term is formally inapplicable.  In particular, we distinguish between topological defects resulting from precise symmetry-breaking conditions classified by non-trivial homotopy groups, and so-called geometric defects which are a consequence of the \textit{topography} of the system. 

Understanding the conditions for existence of topological or geometric defects in a system is essential for understanding its physical properties and potential applications. The occurrence of topological defects in particular can prevent a system from reaching a single domain state, resulting instead in multiple ferroelectric or magnetic domains. These domains are separated by domain walls, which can have novel properties and applications. An example is the conducting  charged ferroelectric domain walls that are trapped in multiferroic hexagonal manganites by virtue of the topology\cite{Meier_et_al:2012}.  In magnetic materials with non-uniform spin textures, the topological ground state is typically more energetically stable than a monodomain phase. As a result it can be controlled at a much lower energy cost, suggesting a plethora of applications in spin-based memory and logic devices\cite{Kiselev_et_al:2011,Fert_et_al:2013}. Geometric defects, on the other hand, are associated with residual entropy which can lead to exotic emergent phenomena such as the monopole-type excitations observed in spin-ice systems\cite{spin-ice}.

\section{Topological defects}

Topological defects were first defined rigorously by Tom Kibble in field theories in the context of cosmological phase transitions in the early universe\cite{Kibble:1976}. The mechanism introduced by Kibble gives the symmetry requirements for the formation of topological defects during a phase transition, and predicts -- based on the particular details of the initial and final symmetries -- what kinds of defects will form. In this section we describe the requirements of the Kibble mechanism, which determine whether a topological defect will form and how it will manifest. We then give three simple examples -- misfit dislocations, skyrmions and ferroelectric domain intersections -- of topological defect formation in materials.

\subsection{Spontaneous symmetry breaking}

The first requirement for topological defect formation within the Kibble mechanism is that the phase transition be spontaneously symmetry breaking. A spontaneously symmetry breaking phase transition is one that exhibits a symmetry change from a higher to a lower symmetry state, which offers multiple degenerate choices of the ground state. (Note that isomorphic phase transitions, such as the evaporation of a liquid to a gas, are not spontaneously symmetry breaking since the symmetry does not change across the transition.) The development of the low symmetry state can be described by the onset of an order parameter, $\phi$, which contains the relevant variables that emerge during the transition. 

A well-known example of a spontaneous symmetry breaking transition is the paramagnetic to ferromagnetic transition in magnets. Here the loss of time-reversal symmetry is described by the onset of a magnetic order parameter, which is often taken to be the magnetization. In the absence of anisotropy, the axis of magnetization in the ferromagnetic state can have any orientation, giving an infinite manifold of degenerate ground states; in real materials, spin-orbit coupling term lifts this rotational symmetry and causes a finite number of preferred easy axes. While many phase transitions in physics are caused by changes in temperature, the concept of spontaneous symmetry breaking is also applicable to changes driven by for example external electric or magnetic fields or pressure.

 \begin{figure}[h]
 \centering
 \includegraphics[width=5cm,keepaspectratio=true]{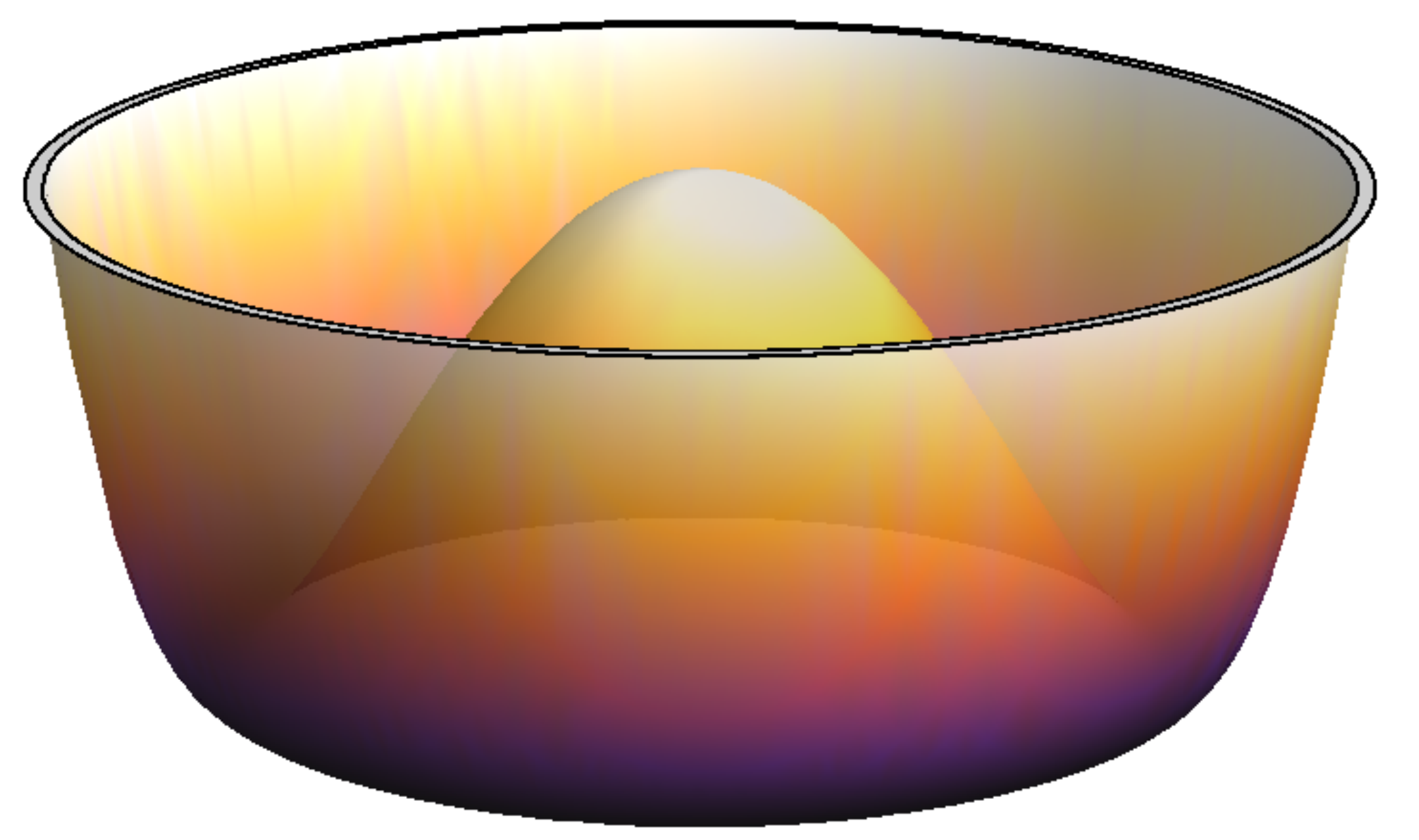}%
 \caption{Mexican-hat potential displaying the degenerate choice of ground states for spontaneous-symmetry breaking phase transitions.}
\label{mexhat}
 \end{figure}
 
The canonical spontaneous symmetry breaking phase transition is described by the `Mexican hat' potential shown in Fig.~\ref{mexhat} with many examples existing in cosmology, high-energy physics and materials science\cite{ssb}. The hat indicates a complex order-parameter function comprising an amplitude and a phase, with the horizontal distance from the peak of the hat giving the magnitude of the order parameter, and the angle around the hat giving its phase. An example of a phase transition in condensed matter described by such a potential is superconductivity, for which the condensate wavefunction is the order parameter. The non-superconducting (high-symmetry) to superconducting (low-symmetry) phase transition spontaneously breaks the symmetry in such a way that the allowed values of the order parameter can point anywhere on a circle (in field-theory language this is described as an $S^{1}$ order-parameter space.) When the system drops from the high-symmetry peak of the hat to the low-symmetry brim, it chooses a particular angle of phase, $\theta$, from the infinite collection of degenerate ground states with $\theta \in \{0, 2\pi\}$, losing its initial $U(1)$ symmetry in the process. Since all choices of $\theta$ have the same energy, there exists a massless Goldstone boson running around the brim of the hat, which is a signature of the initial higher symmetry of the system.

\subsection{Non-trivial homotopy}

Kibble's second condition for topological defect formation is that the symmetry breaking is described by a non-trivial homotopy group\cite{Kibble:1976}. Two topological spaces are homotopic if they can be deformed into each other by continuous transitions such as 	twisting and pulling. The canonical case is the topological equivalence of a coffee cup and a doughnut: A coffee cup can be transformed continuously into a doughnut by pushing the base of the coffee cup upwards until it meets its rim, and then pulling outwards. Thus the transformation between the coffee cup and doughnut can be described by a trivial homotopy group.

In contrast, a bowl and a doughnut are not topologically equivalent. The only way to transform a bowl into a doughnut is to either make a hole in the bowl, or to form it into a long cylinder and attach the ends. Both of these involve cutting and glueing operations, and so the two shapes are homotopically distinct. In this case, the transformation between the bowl and the doughnut is described by a non-trivial homotopy group. 

As a result of the non-trivial homotopy, a phase transition that is described by a non-trivial homotopy group introduces a local kink or crease in the order parameter field on formation of the low-symmetry state. The local kink separates two low-symmetry regions with different values of the order parameter (for example different choices of angle in the Mexican-hat case) and is required to have the structure of the high-symmetry phase in order to resolve the discontinuity in the order parameter. The kink is topologically protected and can only be removed by a global change in the symmetry to the high-symmetry phase.  

\subsection{Classification of topological defects}

A powerful feature of homotopy theory is the ability to classify the topological defects based on the relationship between the initial and final symmetries \cite{Kervaire:1963,Mermin:1979, Nakahara}. This is achieved by mapping the high- and low-symmetry groups onto topological spaces and classifying the relationship between them in terms of its homotopy group. A rigorous mathematical treatment is given in the Appendix -- here we summarize the main results of the relationship between the initial and final symmetry groups and the order parameter field.  We first look at the properties of the order parameter space, and next address the symmetry  change through the phase transition.

Let us begin with a model that is invariant under some symmetry group $G$. If the symmetry group $G$ is spontaneously broken, then there exists a field which remains a solution to the theory, but that is no longer invariant under $G$. We identify this field with the order parameter field and define the parameter space of the ordered, low-symmetry system as $M=G/H$ where $H$ describes the symmetry of the order parameter. Finally, we describe our ordered medium by a mapping, $\phi$, from the real space manifold $A$ into the order parameter space manifold $M$. A stable topological defect occurs when there is a discontinuity in the mapping, $\phi$, between $A$ and $M$, provided that the discontinuity is not caused by pathological behaviour in $\phi$.

The type of defects created are then determined by the order of the homotopy group and can be obtained from standard homotopy tables. The defects that are allowed by homotopy theory in condensed-matter systems are listed for various dimensions in Table ~\ref{defecttable}. Topologically protected surfaces manifest as domain walls in for example ferroelectric or ferroelastic materials. These domain walls separate domains that form in ferroics because $M$ has disconnected components as a result of the discrete symmetry breaking. In cases where $M$ is not simply connected, then unshrinkable loops become trapped around holes in the manifold, leading to the formation of strings such as those formed in liquid Helium and superconductors\cite{Volovik:2003}. Monopoles should form when $M$ contains unshrinkable surfaces, although the elusive magnetic monopole, predicted in Grand Unified Theories, remains to be observed. \footnote{Note that condensed matter analogues of the magnetic monopole such as emergent monopoles in spin ice\cite{spin-ice}, and magnetoelectric monopoles in magnetoelectrics\cite{Spaldin_et_al:2013} are not true monopoles and are not topological defects}. Finally, textures form when more complex symmetries are broken, resulting in delocalized topological defects such as 2D spin textures\cite{Dussaux_et_al:2016}.

\begin{center}
\begin{table}
\caption{Classification of the possible topological defects in condensed-matter systems for $1 \leq$ D $\leq 4$ where D is the dimension of the system.}
\begin{tabular}{|c||c|c|c|c|}
\hline
Homotopy       & D=1 & D=2 & D=3 & D=4 \\ 
Class          &     &     &     &     \\ \hline
$\pi_{0}$ & Monopole & Vortex & Surface & Hypersurface\\ 
$\pi_{1}$ & Texture & Monopole & Vortex & Surface\\
$\pi_{2}$ & - & Texture & Monopole & Vortex\\
$\pi_{3}$ & - & - & Texture & Monopole\\
$\pi_{4}$ & - & - & - & Texture \\ \hline
\end{tabular}
\label{defecttable}
\end{table}

\end{center}

\subsection{Examples of topological defects in condensed matter}
Next we look at three examples of topological defects in condensed matter systems: misfit dislocations, skyrmions, and vortices and domain intersections in multiferroics. 

\subsubsection{Misfit dislocation}
\begin{figure}[h]
 \centering
 \includegraphics[width=8cm,keepaspectratio=true]{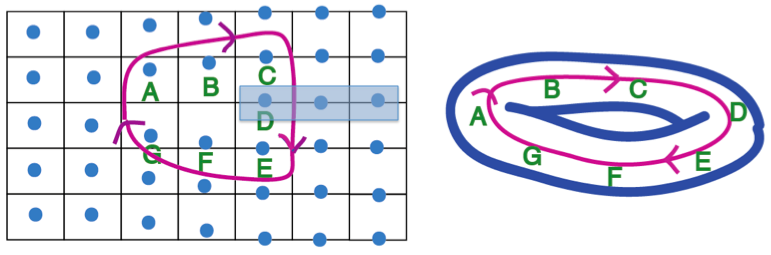}%
 \caption{Loop surrounding a misfit dislocation. The shaded rectangle marks where the extra row appears. The order parameter space is described by a torus (right).}
\label{dislocation}
 \end{figure}

The first example that we discuss is a simple misfit dislocation as shown in Fig.~\ref{dislocation}. The topological defect is located where the extra row disappears; no local rearrangement of the ions will remove the defect. To examine this from a topological perspective, we first identify the order parameter space which must have two translationally symmetric dimensions for the case of a two-dimensional square lattice. In fact the order parameter space is defined by a torus since wrapping a 2D plane in the \textit{x}-direction results in a cylinder and subsequent imposition of boundary conditions in the \textit{y}-direction joins the ends of the cylinder to form a torus.

To analyze the properties of the defect, we circumnavigate the defect noting the motion of the order parameter on the torus. From site \textit{A} to site \textit{B}, we see that the order parameter has an upwards drift with respect to the ideal lattice. Continuing around the defect, there is a continuous drift of the ions with respect to the perfect lattice until point \textit{C} which again coincides with an ideal lattice site but with an extra row of ions in the \textit{y}-direction. In the order parameter space, this gives us a loop around the torus as shown in  Fig.~\ref{dislocation} (right). The torus contains one hole which corresponds to the extra row of ions. In order to return to exactly the same point, the lattice shifts by one complete ion in looping around the defect once, corresponding to a winding number of 1. 

\subsubsection{One-dimensional skyrmion}

 \begin{figure}[h]
 \centering
 \includegraphics[width=7cm,keepaspectratio=true]{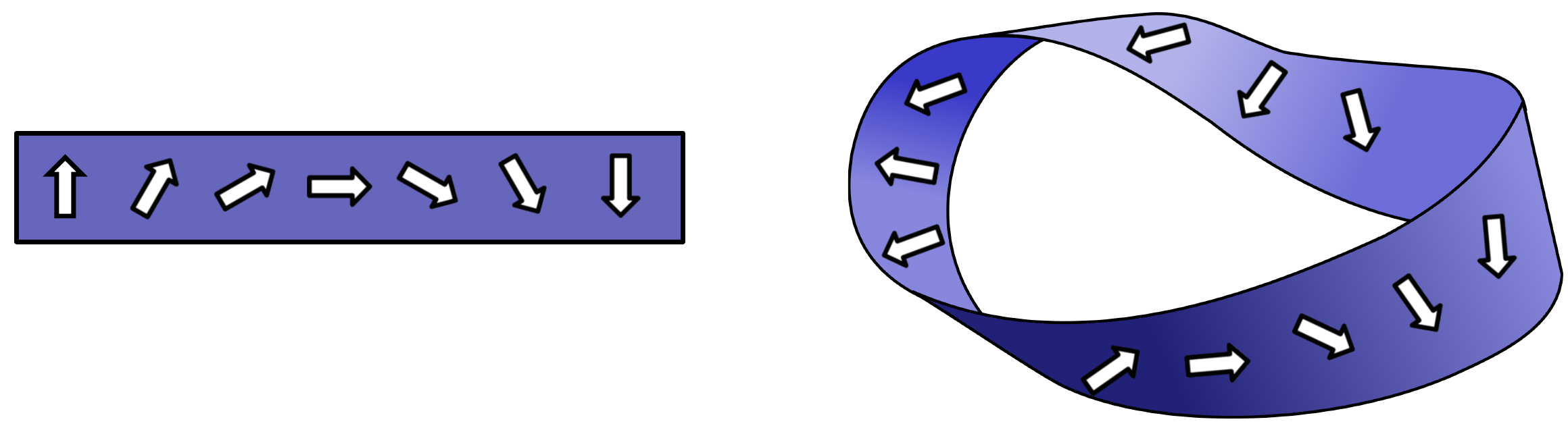}
 \caption{A one-dimensional skyrmion. To restore continuity of the order parameter we introduce a kink in the manifold resulting in a 
 M{\"o}bius strip.}
\label{skyrmion}
 \end{figure}
 
A 1D skyrmion provides our second example of a topological defect\cite{skyrmion}. We consider a 1D lattice composed of magnetic XY spins. In the high-symmetry phase the solution is paramagnetic with disordered spins. At the transition to the skyrmion phase the spins form a spiral as sketched in Fig.~\ref{skyrmion} (left). Imposing conventional periodic boundary conditions to the case shown would cause a discontinuity in the order parameter since it would place an `up' spin directly next to a `down' spin. To remedy this, the order parameter space twists once to restore the continuity of the order parameter and becomes a M{\"o}bius strip (Fig.~\ref{skyrmion} (right)). 

\begin{figure}[h]
 \centering
 \includegraphics[width=7cm,keepaspectratio=true]{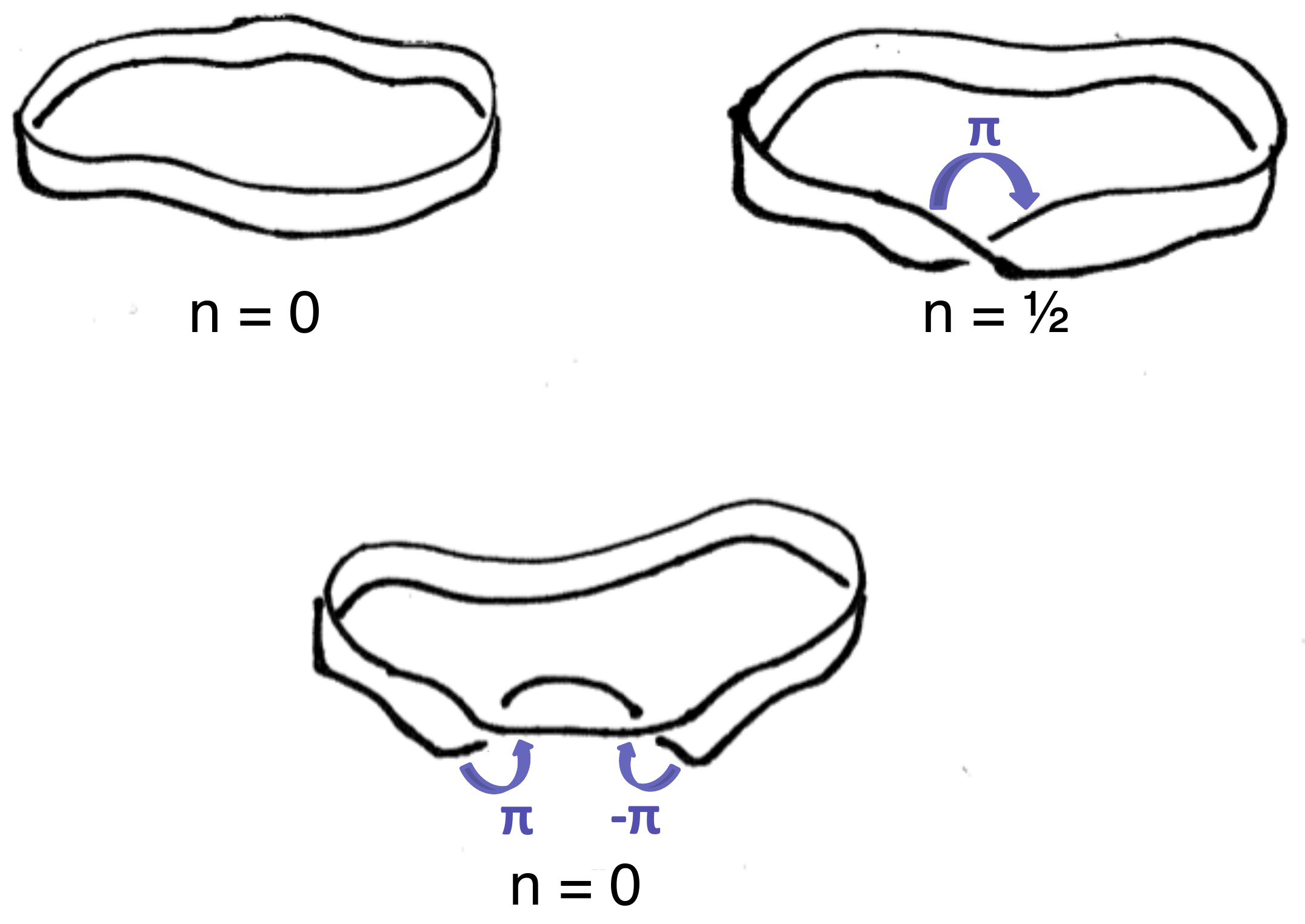}%
 \caption{Loops with different numbers of twists, and their corresponding topological charges, $n$.}
\label{twists}
 \end{figure}

By comparing the properties of a loop with and without a twist, we can determine the topological charge, $n$, carried by the defect in the 1D case, as shown in Fig.~\ref{twists}. A loop with no twists has $n=0$. In our case, we have a single kink with a change of $\pi$ of the order parameter; this situation is defined to have a topological charge of $n=\frac{1}{2}$.  A loop with two opposite twists, with changes of $+\pi$ and $-\pi$ again has  $n=0$.

\subsubsection{Vortices in Multiferroics}

 \begin{figure}[h]
 \centering
 \includegraphics[width=7cm,keepaspectratio=true]{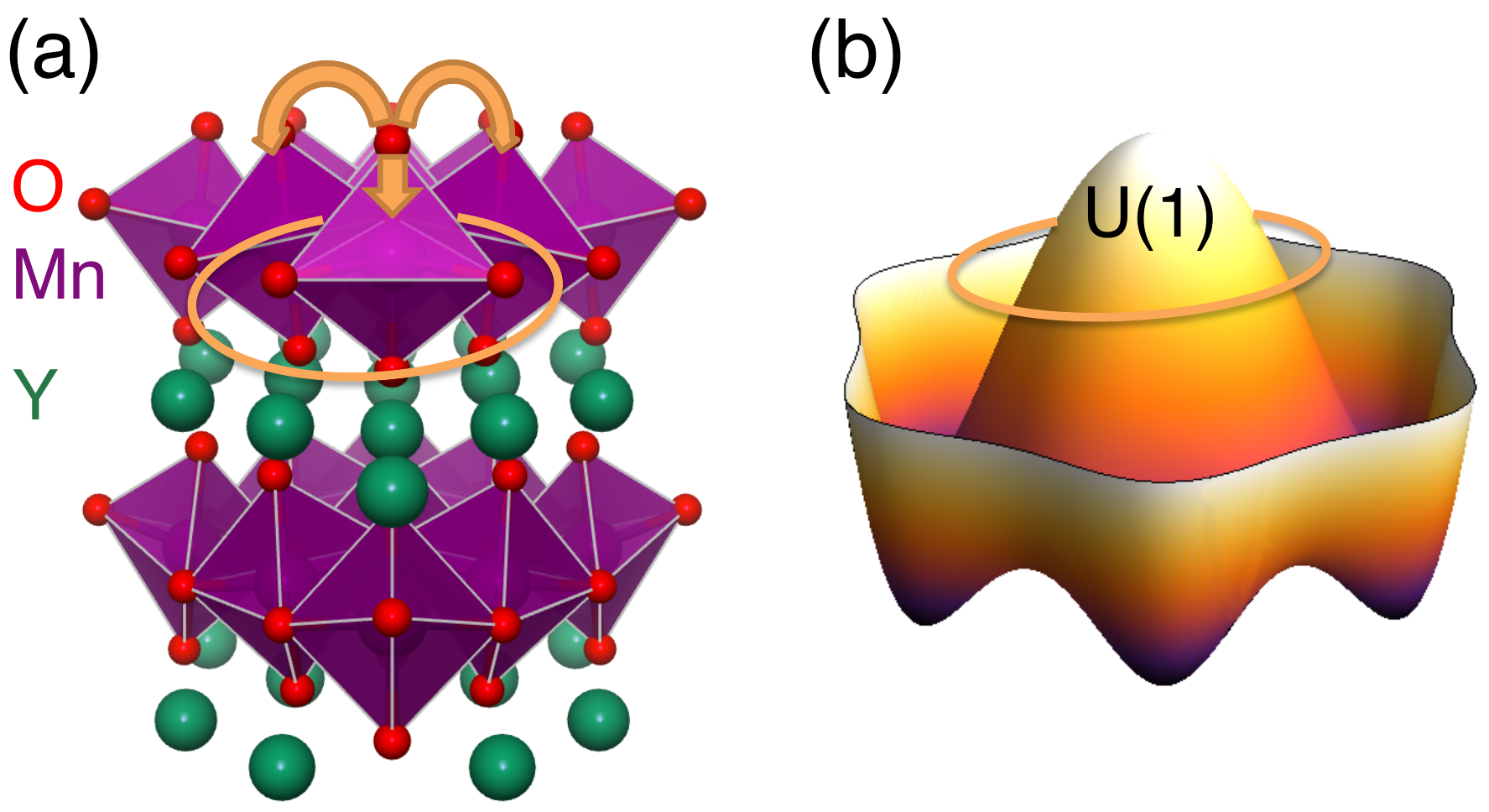}%
 \caption{(a) $R$MnO$_{3}$ structure at the onset of the ferroelectric phase transition. The tilting action of the trimerizing $K_{3}$ mode is shown with yellow arrows and the degenerate 360$^{\circ}$ tilting angles of the MnO$_{5}$ polyhedra is indicated by the yellow circle. (b) Mexican-hat potential energy surface of the hexagonal manganites. At high energy (the peak of the hat) the energy is independent of the angle of trimerization, and the system has $U(1)$ symmetry. At lower energy (in the brim of the hat), six of the trimerization angles become favorable, reducing the symmetry to a six-fold discrete $Z_{6}$ symmetry.}
\label{hexmag}
 \end{figure}
 
Finally we discuss the case of the multiferroic hexagonal manganites $R$MnO$_{3}$ ($R$=Sc, Y, Dy, Ho, Er, Tm, Yb, Lu, In)\cite{Bertaut/Pauthenet/Mercier:1963,Yakel_et_al:1963,Luk74}. These materials are intriguing because they display one-dimensional topologically protected vortices associated with their ferroelectric phase transition\cite{Griffin_et_al:2012}, which occurs $T_{C} \sim 1200$ K, depending on the $R$ cation.

The crystal structure of $R$MnO$_{3}$ consists of planes of corner-sharing MnO$_{5}$ trigonal bipyramids separated by triangular planes of $R$ cations, as shown in Fig.\ref{hexmag}(a)\cite{vanAken_et_al:2004}. The high-temperature phase is centrosymmetric with the $P6_{3}/mmc$ space group. The spontaneous symmetry breaking phase transition at the ferroelectric Curie temperature is unusual in that it is driven by condensation of a trimerizing tilting mode of $K_{3}$ symmetry, with the ferroelectric $\Gamma_{2}^{-}$ mode coupling to it as  a secondary order parameter\cite{Lonkai_et_al:2004,vanAken_et_al:2004,Fennie/Rabe_YMO:2005}. First-principles calculations\cite{Fennie/Rabe_YMO:2005} and Landau theory analysis\cite{Artyukhin_et_al:2014} show that for small amplitudes, the  energy lowering  provided by the condensation of the $K_{3}$ mode is independent  of the tilt angle, as shown in Fig.\ref{hexmag}. The onset of the ferroelectric phase transition can therefore be treated in terms of the breaking of a {\it continuous} $U(1)$ symmetry, in which the full rotational symmetry is broken by tilting of the polyhedra in the full $2\pi$ range of angles. The corresponding potential energy surface resembles a Mexican hat (Fig.\ref{hexmag}(b))\cite{Griffin_et_al:2012}.

The $U(1)$ symmetry of the order parameter space close to the transition allows us to directly apply homotopy theory to predict the resulting topological defects. The order parameter symmetry group, $U(1)$, is first mapped to its corresponding topological space, the one-dimensional circle $S^{1}$. From homotopy tables\cite{Kervaire:1963}, we find that the homotopy group  of this space, $\pi_{k}(S^{1})$ is non-trivial and in fact produces one-dimensional topological singularities -- strings or vortex cores\cite{Kibble:2000}.

With further temperature decrease, the discreteness of the lattice begins to manifest in the energy landscape resulting in a lifting of the 360$^{\circ}$ degeneracy. The polyhedra locking into tilt angles of $0$, $2\pi/3$ or $4\pi/3$, described by $Z_{3}$ symmetry; combined with an additional degeneracy in their tilting direction (`in' or `out') with $Z_{2}$ symmetry, the final order parameter manifold is comprised of six elements and described by $Z_{2} \times Z_{3} \cong Z_{6}$.  This additional coupling to the lattice causes a discrete symmetry breaking which supplements the previous vortex formation with the six domains surrounding each vortex core, separated by six domain walls. Since neighbouring domains have opposite polarization, this allows the direct observation of the domains and hence the topological vortices by piezo-force response microscopy\cite{Choi_et_al:2010,Lilienblum/Soergel/Fiebig:2011}.

\subsection{Summary}
To summarize this section, topological defects are features that result from strict symmetry requirements associated with certain symmetry-breaking phase transitions. The Kibble mechanism gives the requirements for their formation using mathematical homotopy theory. First, the symmetry must be spontaneously broken, that is, that the ground state is degenerate. The second requirement is that the symmetry change corresponds to a non-trivial homotopy group. The nature of the homotopy group also predicts the type and dimension of the resulting topological defect.  Dislocations, skyrmions and ferroelectric domain intersections  are examples of topological defects in materials.

\section{Geometric defects}

Geometric defects occur when perfect long-range order is prohibited from forming because of short-range constraints, which can result from the connectivity of the lattice, the magnetic interactions or the details of the chemical bonding. Such systems are often referred to as geometrically frustrated. Legend has it that the term `frustration' was first used in 1976 by P. W. Anderson who wrote ``Frustration is the name of the game'' on an Aspen blackboard\cite{Buschow:2003}; the term was first seen in the literature in works by Toulouse\cite{Toulouse:1977} and Villain\cite{Villain:1988}. Here we discuss three examples of geometrical frustration in solid state physics: The  tiling of a plane with regular pentagons, which is relevant in the analysis of quasicrystals;  antiferromagnetic interactions between magnetic moments on a triangular or tetrahedral lattice, such as occur in multiferroic hexagonal manganites and Kagome lattices; and finally water- and spin-ice materials, such as the rare-earth pyrochlores, which have a residual entropy and several interesting emergent properties.

\subsection{Examples of geometric defects in condensed matter}
\subsubsection{Pentagonal tiling and quasicrystals}

 \begin{figure}[h]
 \centering
 \includegraphics[width=8cm,keepaspectratio=true]{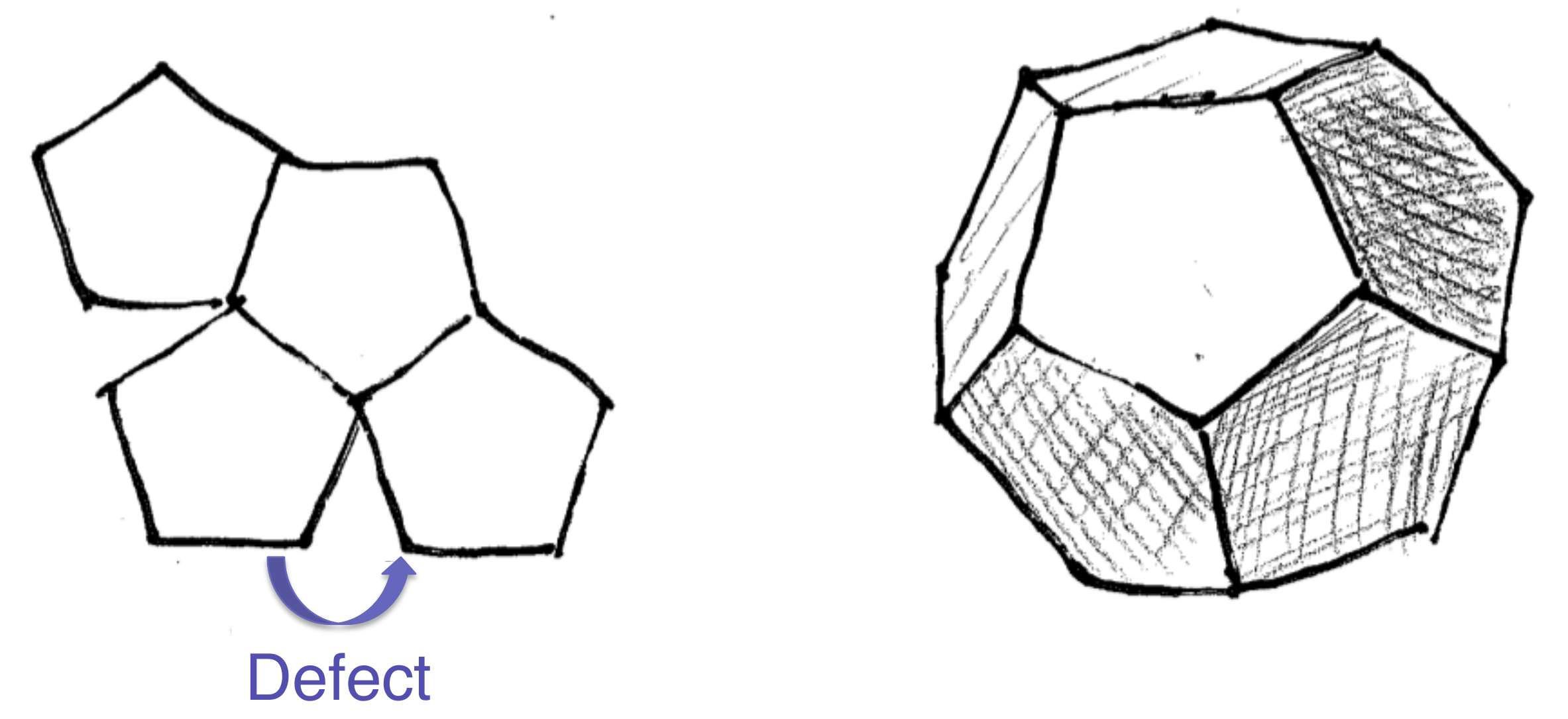}%
 \caption{Pentagons cannot tile a 2D plane. The defects are removed when applied to a higher-dimensional space,
 forming a dodecahedron.}
\label{pentagon}
 \end{figure}
 
We begin by trying to tile a 2D plane with regular polygons, as shown in Fig.~\ref{pentagon}. While squares or hexagons can easily be arranged to completely fill a 2D plane, tiling a plane with regular pentagons is impossible without forming kinks called defects (when a gap remains between adjacent pentagons) or excesses (when adjacent pentagons overlap). These are geometric defects, and are a result of the incompatibility of the local-ordering rule of regular pentagons with the 2D surface that it is tiling. 

The local-ordering rule of regular pentagons is compatible, however, with tiling a 3D (or higher-dimensional) surface. In dimensions greater than 2, the curvature of the space allows the pentagons to fit neatly together avoiding the formation of geometric defects and forming a 3D ``football'', as shown in Fig.~\ref{pentagon}. 

This simple example illustrates the two most relevant characteristics of geometric defects: First, a local constraint (regular pentagons) prevents long-range ordering from forming with the production of geometric defects. Second, to achieve long-range order, one must change the global space on which the local ordering is applied, in this case increasing the dimension of the tiling plane from two to three.

The analogous exercise in three dimensions leads to the well-known rule that a crystal can not have five-fold rotational symmetry.  Indeed, the report in 1984 of an apparent five-fold symmetry in the Bragg diffraction from alumunim-manganese alloys \cite{Schechtman_et_al:1984} was so transformative in our understanding of crystallography that it was awarded the 2011 Nobel prize in Chemistry. We now call materials showing this behavior \textit{quasicrystals}, to reflect the fact that they have long-range order but no long-range translational symmetry. 

\subsubsection{Triangular antiferromagnetism} 

\begin{figure}
 \centering
 \includegraphics[width=7cm,keepaspectratio=true]{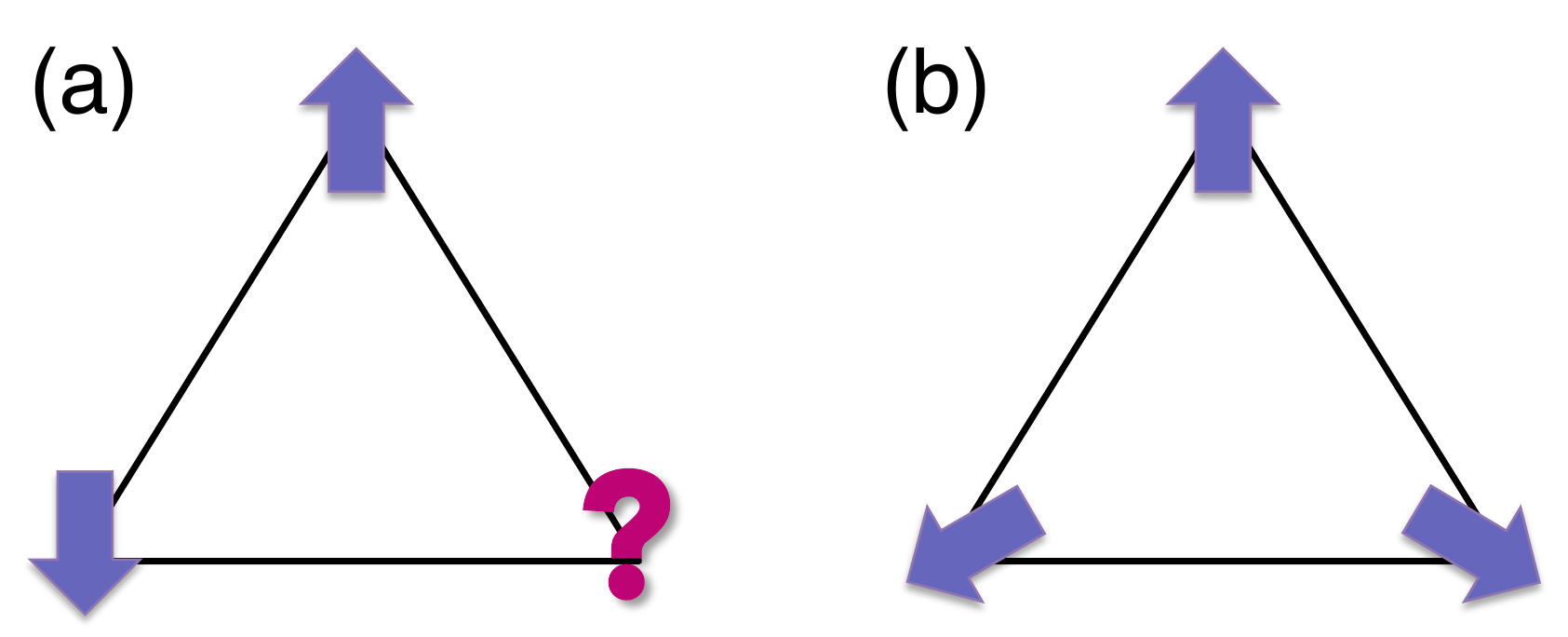}%
 \caption{(a) Collinear Ising spins on a triangular lattice. If the spins have antiferromagnetic coupling, they cannot form a long-range ordering. (b) Noncollinear spins on a triangular lattice. The antiferromagnetic coupling between the spins drives formation of a 120$^{\circ}$ spin structure.}
\label{ising}
 \end{figure}

Another example of geometric defects occurs in the case of antiferromagnetically coupled collinear magnetic moments (Ising spins) on a triangular lattice. This scenario was first studied by Wannier\cite{Wannier:1950} and is depicted in Fig.~\ref{ising}. Once the two left-most spins have coupled antiferromagnetically, the total energy is independent of the orientation of the remaining spin and the system is frustrated. As a result long-range order is prevented, and  a measurable entropy results from the large number of possible geometrically-frustrated arrangements.

In Fig.~\ref{ising} we show a route to achieving compatibility between the local ordering rules and the global space. By moving away from the simple Ising picture, which only allows for collinear spins, the local ordering can adopt the triangular compromise shown. This removes both the frustration and the defects.  The order parameter space now allows more solutions than simply `up' and 'down' in the Ising case. 

Interestingly, noncollinear triangular antiferromagnetism is observed in the hexagonal rare earth manganites described above\cite{Bertaut:1963,Fiebig_et_al:2000}. The Mn atoms form a triangular sublattice, and below T$_{N}$~$\sim$ 100 K their magnetic moments couple antiferromagnetically via the superexchange interaction. Because of the incompatibility between the lattice structure and the exchange coupling, a noncollinear 120$^{\circ}$ magnetic ordering is the ground state.

\subsubsection{Water ice and spin ice}

 \begin{figure}[h]
 \centering
 \includegraphics[width=7cm,keepaspectratio=true]{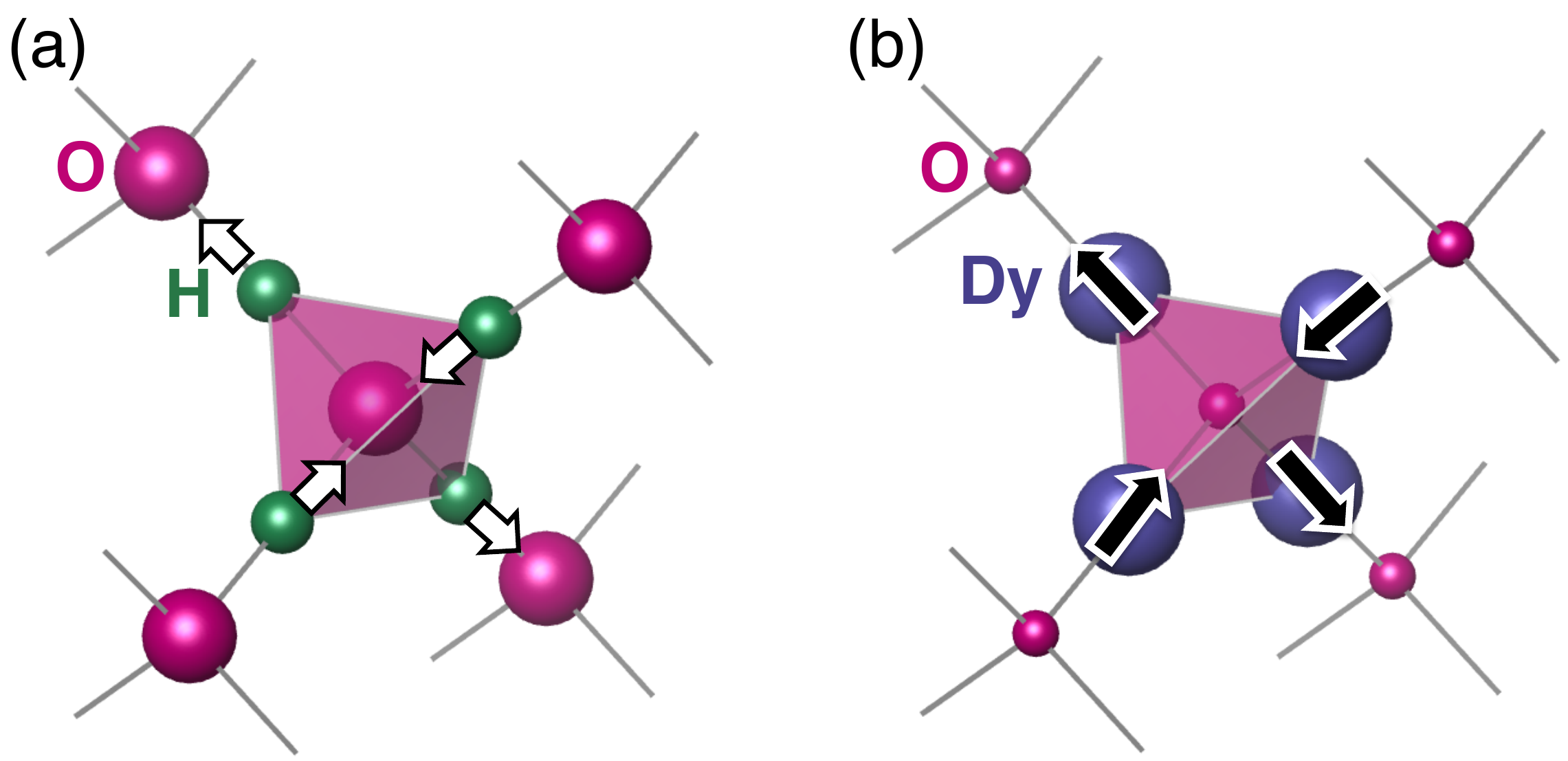}%
 \caption{(a) Water ice in the I$_{h}$ structure with the `two-in', `two-out' arrangement of H atoms surrounding an O atom. (b) Spin-ice Dy$_{2}$Ti$_{2}$O$_{7}$ with the Ising spins of the Dy atoms pointing in a `two-in', `two-out' configuration along the tetrahedral axes.}
\label{ice}
 \end{figure}

In 1936, calorimetry measurements on water ice down to 15 K revealed a residual entropy, that is an entropy greater than that expected for a crystalline state \cite{Giauque/Stout:1936}.  Linus Pauling  explained this by considering configurational disorder in the local arrangements of the  hydrogen atoms surrounding each oxygen\cite{Pauling:1935}, using the Bernal-Fowler `ice rules'\cite{Bernal/Fowler:1933}. In the cubic and hexagonal forms of ice, each oxygen atom  is tetrahedrally coordinated by four hydrogen atoms, while each hydrogen has two oxygen neighbors (Fig.\ref{ice}(a)). In order to maintain local H$_{2}$O entities, each oxygen forms strong, short covalent bonds with two neighboring hydrogens, and weak, long hydrogen bonds with the other two neighboring hydrogens. The very large number of possible `two-in', `two-out' arrangements is the origin of the residual entropy. In this case the geometric frustration arises from the incompatibility between the tetrahedral hydrogen-oxygen network and the local bonding rules of the chemical environment.  

An analogous situation occurs in the pyrochlore compounds Dy$_{2}$Ti$_{2}$O$_{7}$ and  Ho$_{2}$Ti$_{2}$O$_{7}$, commonly referred to as `spin ice'\cite{spin-ice}. In the pyrochlore lattice, the rare-earth ions Dy and Ho occupy the corners of tetrahedra. Because of the strong spin-orbit coupling of these elements, their spins are Ising-like with their easy axes along the tetrahedral axes connecting the center of the tetrahedra to the vertices (Fig.\ref{ice}(b)).  The minimal energy solution is for two spins to point towards the center of the tetrahedron, and two to point outwards, resulting in the `two-in', `two-out'  spin-ice rules. Like the water-ice case, the spin-ice materials have non-zero residual entropy\cite{Ramirez_et_al:1999}, and furthermore host fascinating emergent properties such as monopoles and Dirac strings as a result of the geometric frustration\cite{Castelnovo_et_al:2008}.

Note that in these examples the new global space that allows restoration of the local order has the same \textit{topology} as the initial space and so the defects are not topological.

\section{Summary}

We have reviewed the concepts of topological defects and geometric defects, using simple examples from materials and condensed-matter physics to illustrate their occurrence and properties. 

An obvious connection exists between topological and geometric phenomena in the sense that both are caused by symmetry constraints, and both have a feature that precludes the system from reaching a homogeneous global minimum. In the case of topological defect formation, the order parameter describing the phase transition has a `kink' which can never be ironed out, whereas geometric frustration prevents the system from adopting long-range order through a kink in the short-range order.

The two defect types have important differences, however. While topological defects result from an incompatibility between the topology of the initial and final symmetry groups across a phase transition, geometric defects are caused by an incompatibility between local ordering rules and the space or lattice in which they are applied. Topological defects can only be removed by changing the topology of one of the underlying spaces, for example by the introduction of holes or twists, so that the initial and final symmetry groups have the same homotopy. In contrast, geometric defects can be removed by changing the global symmetry to that of a suitable higher symmetry group allowing restoration of the local ordering.

Both types of defects are associated with myriad fascinating material properties, many of which have only recently been observed. We expect that there are many more awaiting discovery.

\section{Acknowledgments}
We thank Ekhard Salje for inspiring and informative discussions regarding the formation and properties of domain walls and vortex intersections. We also thank Andrea Scaramucci and Tess Smidt for insightful feedback and comments. This work was supported by the ERC Advanced Grant program, No. 291151, by the Max R\"ossler Prize of the ETH Z\"urich and by the Koerber Foundation. 

\section*{Appendix: Mathematical description of Kibble mechanism for defect formation}

The first homotopy group $\pi_{1}(M)$ is also called the fundamental group of the manifold $M$ at $a$ because all groups $\pi_{1}(M,a)$ are identical.

To generalize the first homotopy group to higher homotopies, a fundamental theorem will be stated (without proof) and then applied to a symmetry-breaking process. The main point is that to determine the homotopy group of a symmetry breaking, we need not know both the high- and low-symmetry groups $G$ and $H$. Rather it is sufficient to know the topology of the vacuum manifold, $M = G/H$. 

Let a simply-connected\footnote{The restriction of simply-connected group can be imposed by extending any (compact, Lie) group into a universal covering group.} $G$ be spontaneously broken to a subgroup $H$. The vacuum manifold $M$ is then given by the space of cosets $H \subset G$, $M = G/H$.

Let the subgroup $H$ have a component $H_{0}$ connected to the identity. The disconnected components of $H$ can be labelled by the quotient group $\pi_{0}(H)=H/H_{0}$. This is isomorphic to the fundamental group of the coset space $\pi_{1}(G/H)$, that is
\begin{equation}
\pi_{1}(G/H) \cong \pi_{0}(H)
\end{equation}

Thus, whether or not defects will form is determined by the topology of the vacuum manifold. Consider the order parameter mapping $\phi: A \rightarrow M$. The defect condition is then
\begin{equation}
\pi_{k}(M) \neq 1
\end{equation}
where $\pi_{k}(M)$ is the \textit{k$^{th}$}-homotopy group of $M$ and $k \in \mathbb{R}$. To find the homotopy group $\pi_{k}(M)$, the various classes of homotopy mappings from $S^{k}$ to $M$ are found. If $M$ is some \textit{m}-sphere, the  $\pi_{k}(S^{k})$ is only non-trivial when $k \geq m > 1$. Considering the case for $m=1$, the $k^{th}$ homotopy $\pi_{k}(S^{k})$ is only non-trivial when $k=1$. To classify defects, the unbroken symmetry group is all that is necessary, provided we started with a simply-connected group in the beginning. So in order to have a defect, a non-trivial group is needed in the first place.

\bibliography{Nicola,sinead}

\end{document}